\documentclass[conference]{IEEEtran}
\IEEEoverridecommandlockouts
\usepackage{cite}
\usepackage{amsmath,amssymb,amsfonts}
\usepackage{algorithmic}
\usepackage{graphicx}
\usepackage{textcomp}
\usepackage{xcolor}
\usepackage{todonotes}
\usepackage{hyperref}
\usepackage{cleveref}
\usepackage{makecell}
\def\BibTeX{{\rm B\kern-.05em{\sc i\kern-.025em b}\kern-.08em
    T\kern-.1667em\lower.7ex\hbox{E}\kern-.125emX}}

\crefname{figure}{Fig.}{Figs.}
\Crefname{figure}{Fig.}{Figs.}
\crefname{table}{Table}{Tables}
\Crefname{table}{Table}{Tables}
\crefname{equation}{}{}
\Crefname{equation}{Equation}{Equations}
\crefname{section}{}{}
\Crefname{section}{Section}{Sections}
\begin{document}

\title{Benchmarking Quantum Annealers with Near-Optimal Minor-Embedded Instances}

\author{\IEEEauthorblockN{Valentin Gilbert}
\IEEEauthorblockA{\textit{Université Paris-Saclay} \\
\textit{CEA, List, F-91120}\\
Palaiseau, France \\
valentin.gilbert@cea.fr}
\and
\IEEEauthorblockN{Julien Rodriguez}
\IEEEauthorblockA{\textit{Université de Montpellier} \\
\textit{LIRMM, CNRS }\\
Montpellier, France \\
julien.rodriguez@lirmm.fr}
\and
\IEEEauthorblockN{Stéphane Louise}
\IEEEauthorblockA{\textit{Université Paris-Saclay} \\
\textit{CEA, List, F-91120}\\
Palaiseau, France \\
stephane.louise@cea.fr}
}

\maketitle

\begin{abstract}
Benchmarking Quantum Process Units (QPU) at an application level usually requires considering the whole programming stack of the quantum computer. One critical task is the minor-embedding (resp. transpilation) step, which involves space-time overheads for annealing-based (resp. gate-based) quantum computers. This paper establishes a new protocol to generate graph instances with their associated near-optimal minor-embedding mappings to D-Wave Quantum Annealers (QA). This set of favorable mappings is used to generate a wide diversity of optimization problem instances. We use this method to benchmark QA on large instances of unconstrained and constrained optimization problems and compare the performance of the QPU with efficient classical solvers. The benchmark aims to evaluate and quantify the key characteristics of instances that could benefit from the use of a quantum computer. In this context, existing QA seem best suited for unconstrained problems on instances with densities less than $10\%$. For constrained problems, the penalty terms used to encode the hard constraints restrict the performance of QA and suggest that these QPU will be less efficient on these problems of comparable size.

\end{abstract}


\section{Introduction}

The past five years have seen the rise of quantum computers with successfully implemented quantum supremacy experiments. These experiments were demonstrated with superconducting qubits \cite{Arute2019} and with photonic qubits \cite{zhong2020quantum, Madsen2022}. More recently, D-Wave claimed that their QA reached quantum supremacy \cite{king2024computational}. These demonstrations constitute strong experimental steps towards achieving a useful quantum advantage but are insufficient to demonstrate the utility of quantum computers to solve real-world and useful problems. Combinatorial optimization problems constitute a set of relevant problems to evaluate the potential quantum advantage that could be brought by Noisy Intermediate Scale Quantum (NISQ) process units. The Q-Score is an example of application-oriented protocol designed to benchmark QPU using random max-cut instances based on Erdős-Rényi graphs\cite{martiel2021benchmarking}. Several frameworks have been recently designed to benchmark annealing-based and gate-based models in a holistic way \cite{mills2021application}. One example is the QUARK framework, which explores real industrial use cases \cite{finvzgar2022quark}. Various national and international initiatives are developing application-oriented frameworks, including the QED Consortium \cite{lubinski2023application} and the BACQ project\cite{barbaresco2024bacq}. These frameworks are completed with attempts to quantify the coverage of instance sets to gauge the fairness of a benchmark experiment \cite{tomesh2022supermarq, Gilbert2023}. Recent studies gather specific instance sets designed to benchmark quantum computers, such as HamLib \cite{sawaya2023hamlib}, which proposes various encodings of Hamiltonians for optimization problems and condensed matter models. The MQTBench library \cite{Quetschlich2023} establishes a database of quantum circuits that can be used for gate-based QPU. One can also rely on previously developed reference libraries of problems used for classical computers, such as the MQLib \cite{Dunning2018} and the QPLib \cite{Furini2018}. The interested reader can refer to a recent survey done on this subject \cite{abbas2023quantum}.

The methods described above are based on the same execution scheme: an input instance is defined and a preprocessing routine is used to map the problem on the QPU. This task is called the compilation or transpilation for gate-based quantum computers. It converts the original gate set to a physically realizable gate set and adds SWAP gates to compensate for the limited interconnects of the quantum chip. For QA such as D-Wave Systems \cite{DWaveInternalDocumentation}, the preprocessing step consists of finding a transformation that maps each problem's variable to a set of physical variables that can be straightly encoded to the physical qubits of the quantum annealer. This task is called minor-embedding \cite{choi2008minor}. Transpilation as well as minor-embedding tasks are hard for classical computers and directly impact the quality of the solutions provided by the QPU.

A second approach uses sets of \textit{crafted instances} to benchmark the QPU \cite{abbas2023quantum}. These instances are designed to reduce the impact of the preprocessing step and constitute friendly use cases for the QPU but remain hard for classical optimizers. \Cref{tab:crafted_instances} shows recent experiments led on D-Wave QA using this type of instance. Each experiment shows that QA can excel at solving specific sets of instances compared to classical solvers. These instances have two common features: they have many variables and are almost always subgraphs of the qubit layout, which avoids using extra qubits to embed the problem on the quantum chip. \textit{Crafted instances} have been used to compare the performance of IBM gate-based QPU to D-Wave QA, leading to the superiority of QA for solving optimization problems \cite{pelofske2023quantum}.

\setlength\tabcolsep{4pt}
    \begin{table*}[t]
      \centering
      \caption{Related work using crafted instance sets to demonstrate the performance of D-Wave quantum annealers.}
      \begin{tabular}{|c|c|c|c|c|c|c|c|}
        \hline
            Source& Instance type & Instance topology & \#Variables & \#Qubits  & Tested QPU & Minor-Embedding  & Classical solvers  \\
            \hline
            \cite{pang2021potential} (2021) & BFM$^{\mathrm{*}}$, FBFM$^{\mathrm{*}}$, CBFM$^{\mathrm{*}}$ & Chimera graph & 2032  & 2032 &  \textit{2000Q} & QPU chip sub-graph & Global \& local search \\
            \hline
            \cite{tasseff2022emerging} (2022) &  CBFM$^{\mathrm{*}}$-P & Pegasus graph & 5387 & 5387 &  \textit{Advantage\_System4.1} & QPU chip sub-graph & Global \& local search \\
            \hline
            \cite{lubinski2023optimization} (2023) & Unweighted Max-cut & 3-regular graph &  4-320 & 4-320 & \textit{Advantage\_System4.1} & QPU chip sub-graph & Exact solver \\
            \hline
            \cite{king2024computational} (2024) & \makecell{2D \& 3D Ising \\ spin glass} & \makecell{ square, cubic, \\ diamond, biclique } & 16-567 & 16-576 & \makecell{\textit{Advantage\_System4.1} \\ and \textit{Advantage2}}  & \makecell{QPU chip sub-graph \\ 2 qubits/var (biclique)} & Exact solver \\
        \hline
        \multicolumn{8}{l}{$^{\mathrm{*}}$BFM, FBFM and CBFM respectively relates to Biased Ferromagnet, Frustrated Biased Ferromagnet and Corrupted Biased Ferromagnet Ising models.}
      \end{tabular}
      
      \label{tab:crafted_instances}
\end{table*}

This paper provides a new method to create large sets of \textit{crafted instances} that are near-optimally mapped on a given QPU. Our study mainly focuses on QA, which are the most mature quantum technologies able to approximate large optimization problems. However, our method can be extended to benchmark gate-based QPU. The creation of the instance set is designed in a reverse fashion: At first, a near-optimal mapping of a complete graph (source graph) on the quantum chip (target graph) is found. The mapping function is then iteratively altered to increase the size of the source graph while decreasing its density. During this process, the target graph is not altered. The set of source graphs is then used to generate instances of optimization problems with various densities to benchmark the performance of QA. Our analysis provides valuable insights into the characteristics of instances that could potentially lead to a quantum advantage. Our finding is that despite near-optimal embeddings, the first generation of D-Wave Advantage rapidly struggles to find good solutions for instances with densities greater than 0.1.

The remainder of the paper is organized as follows: \Cref{section:2} introduces the basic notions and notations on quantum annealing and minor-embedding. \Cref{section:3} presents the method used to create the \textit{crafted instances} and the technical settings of the experiment. \Cref{section:4} benchmarks the QPU against classical solvers and discusses the results.

\section{Notation and definitions}\label{section:2}


Quantum annealers, such as D-Wave systems, constitute a noisy version of the universal adiabatic quantum computer \cite{albash2018adiabatic}. The evolution of these systems is based on the interpolation of a Mixing Hamiltonian $H_\mathrm{M}$ whose ground state is easy to prepare and a problem Hamiltonian $H_\mathrm{P}$ whose ground state encodes the solution to the problem. The problem Hamiltonian can be fully specified with a source graph $G_\mathrm{s}=(V_\mathrm{s}, E_\mathrm{s})$:
\begin{equation}
    H_\mathrm{P} = \sum_{v \in V_\mathrm{s}} h_v \sigma_v^z + \sum_{(u, v) \in E_\mathrm{s}} J_{uv} \sigma_u^z \sigma_v^z
\end{equation}
where $\sigma_i^z$ denotes the Pauli Z operator on variable $i$. The ground state of $H_\mathrm{p}$ encodes the solution to the problem:
\begin{equation}
    \min C(\mathbf{s}) = \sum_{v \in V_\mathrm{s}} h_v s_v + \sum_{(u, v) \in E_\mathrm{s}} J_{uv} s_u s_v
    \label{eqn:ising_cost_function}
\end{equation}
where $s_i \in \{-1, +1\}$. A Quadratic Unconstrained Binary Optimization (QUBO) cost function with variables $x_i \in \{0,1\}$ is translated to an Ising cost function \cref{eqn:ising_cost_function} using a variable substitution $x_i = \frac{1+s_i}{2}$. When a graph $G_\mathrm{s}$ cannot be straightly mapped to the qubits of the quantum chip due to connectivity limitations, one has to find a mapping function that maps the source graph $G_\mathrm{s}$ to a target graph $G_\mathrm{t}=(V_\mathrm{t}, E_\mathrm{t})$ where $G_\mathrm{t}$ is defined as a subgraph of the quantum chip's graph $G_\mathrm{t\_QPU}$. This problem is well-defined in the theory of graph minors \cite{Robertson1995} and is formally stated as:

\textit{Find a function \(\phi : V_\mathrm{s} \xrightarrow{} \mathcal{P}(V_\mathrm{t})\) such that : 
\begin{enumerate}
    \item each vertex \(v \in V_\mathrm{s}\) is mapped onto a connected subgraph \(\phi(v)\) of \(G_\mathrm{t}\). \(G_\mathrm{t}\) is a subgraph of $G_\mathrm{t\_QPU}$.
    \item each connected subgraph must be vertex disjoint \(\phi(v) \cap \phi(v') = \emptyset\), with \(v \neq v'\).
    \item each edge \((u,v) \in E_\mathrm{s}\) is mapped onto at least one edge in \(E_\mathrm{t}\) : \(\forall (u,v) \in E_\mathrm{s}, \exists u' \in \phi(u), \exists v' \in \phi(v),~such~that~(u',v') \in E_\mathrm{t}\).
\end{enumerate}
}
\noindent
Let $n_\phi$ be the number of nodes used in $G_\mathrm{t}$ to embed the graph $G_\mathrm{s}$. A coupling strength is set to each edge of each connected subgraph $\phi(v)$ to maintain ferromagnetic coupling between the physical qubits, which act as a single logical variable. In this paper, we refer to $v \in V_s$ as logical qubits or logical nodes (equivalent to logical variables). We refer to $v \in V_t$ as physical qubits or physical nodes. Each logical node is mapped to a set of physical nodes $\phi(v)$.

Two different types of algorithms are mainly used to find the function $\phi$. The first set of algorithms takes as input both $G_\mathrm{s}$ and $G_\mathrm{t\_QPU}$. One implementation is the CMR heuristic \cite{cai2014practical}, which has the advantage of working with any graph $G_\mathrm{t\_QPU}$. The second set of algorithms take as input $G_\mathrm{s}$ and are designed for specific $G_\mathrm{t\_QPU}$. These algorithms usually take advantage of the structure of $G_\mathrm{t\_QPU}$ and produce mappings that are near-optimal for complete graphs. For example, the CME heuristic \cite{boothby2016fast} finds high-quality mappings of complete graphs for D-Wave quantum chips, leveraging the regular structure of the chip and considering inoperable qubits.

\section{Method}\label{section:3}

\subsection{Assessing the Quality of an Embedding}

The minimum number of physical nodes required to embed a source graph $G_\mathrm{s}$ can be lower bounded. Even if not reachable in practice, this lower bound gives an idea of the quality of the embedding found and the distance to an optimal embedding. Let $ n_{\phi(v)^*}$ be a lower bound on the optimal number of nodes required to embed a node $v \in V_\mathrm{s}$ on the target graph. We assume $G_\mathrm{t\_QPU}$ has a regular topology and define $c_\mathrm{phys}$ as the number of edges per node, which is constant for D-Wave topologies, \emph{e.g.}, $c_\mathrm{phys} = 15$ for Pegasus topology. If $\phi(v)^*$ maps $v \in V_\mathrm{s}$ to a path (structure that maximizes the potential connectivity of $\phi(v)^*$), the node $v$ requires at least $ n_{\phi(v)^*}$ nodes to be embedded on the target graph $G_\mathrm{t}$:

\begin{equation}
   n_{\phi(v)^*} = \left.
  \begin{cases}
    1 \text{ if } \mathit{deg}(v) \leq c_\mathrm{phys} \\
    2 \text{ if } c_\mathrm{phys} < \mathit{deg}(v) \leq (2 c_\mathrm{phys} -2) \\
    \left\lceil \frac{\mathit{deg}(v)-(2 c_\mathrm{phys} - 2)}{c_\mathrm{phys}-2} \right\rceil + 2 \text{ otherwise }
  \end{cases}
  \right\}.
  \label{eqn:opt_phys_qubits}
\end{equation}
The number of nodes used in the target graph to embed $G_\mathrm{s}$ is then $n_{\phi} \geq \sum_{v \in V_\mathrm{s}}  n_{\phi(v)^*}$. This bound is then used to define an overhead ratio $r_\mathrm{o}$ that computes the overhead of physical qubits used by the mapping function:

\begin{equation}
    r_\mathrm{o} = \frac{n_\phi}{\sum_{v \in V_s}  n_{\phi(v)^*}}.
    \label{eqn:overhead_ratio}
\end{equation}
The closer the ratio $r_\mathrm{o}$ is to 1, the better. We use this metric to show that our instance generation process creates near-optimally embedded instances.

\subsection{Near-optimal Mapping Generation}

\begin{figure}[t]
    \centerline{\includegraphics[page=1,width=\columnwidth]{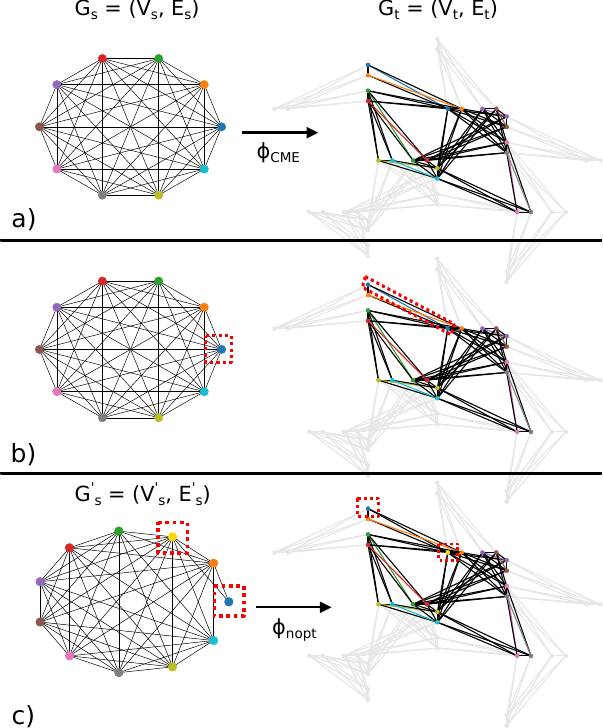}}
    \caption{Instance generation with iterative chain split. a) The algorithm starts with a complete graph embedding using CME method on the graph $G_\mathrm{t\_QPU}$. Each node in the source graph $G_\mathrm{s}$ is mapped to a chain in the target graph $G_\mathrm{t}$. b) A random logical node $v \in V_\mathrm{s}$ is selected. c) The corresponding chain $\phi_\mathrm{CME}(v)$ is split in two parts to create a new logical node in the source graph. It changes the mapping function to $\phi_\mathrm{nopt}$. Steps b. and c. are repeated on $G_\mathrm{s}'$ until the desired density in the logical graph is reached.}
    \label{fig:instance_generation}
\end{figure}

\Cref{fig:instance_generation} shows the workflow creating graph instances near-optimally embedded on the D-Wave quantum chip. The first step consists of finding the mapping function of a complete graph $G_\mathrm{s}$ to a target graph $G_\mathrm{t}$ that is a subgraph of $G_\mathrm{t\_QPU}$. The CME algorithm finds such mapping (see \cref{fig:instance_generation}a). Each logical node $v \in V_\mathrm{s}$ is mapped to a connected subgraph $\phi_{\mathrm{CME}}(v)$ on the D-Wave chip, which topology is represented by the graph $G_\mathrm{t\_QPU}$. One special feature is that the CME algorithm generates subgraphs $\phi_{\mathrm{CME}}(v)$ that are paths to maximize the potential connectivity of each logical qubit $\phi_{\mathrm{CME}}(v)$. It is then possible to increase the size of the logical graph $G_\mathrm{s}$ by selecting and splitting a random path $\phi_{\mathrm{CME}}(v)$ in two equal parts (see \cref{fig:instance_generation}b and c). The new source graph $G_\mathrm{s}' = (V_\mathrm{s}', E_\mathrm{s}')$ has a lower density than $G_\mathrm{s}$, but an additional node owing to the split. This last step is repeated to create arbitrarily dense source graphs $G_\mathrm{s}'$, preserving the near-optimal embedding clause in the new mapping function $\phi_\mathrm{nopt}(v)$. As the number of iterations increases, the Graph Edit Distance (GED) between $G_\mathrm{s}'$ and $G_\mathrm{t}$ shrinks, meaning that the topology of $G_\mathrm{s}'$ becomes closer to the topology of $G_\mathrm{t}$. It is important to notice that the set of nodes $V_\mathrm{t}$ of $G_\mathrm{t}$ are defined by the CME method at the beginning and remain fixed during the steps used to generate the different instances $G_\mathrm{s}'$. Hence, all the instances $G_\mathrm{s}'$ use the same physical graph $G_\mathrm{t}$. Only the mapping function $\phi_\mathrm{nopt}$ varies between each graph $G_\mathrm{s}'$. The generated graphs $G_\mathrm{s}'$ that are very sparse are considered favorable to the quantum computer as they have many logical variables, each mapped to very few or only one physical qubit. On the contrary, graphs $G_\mathrm{s}'$ with high densities have each vertex $v$ mapped to long chains of physical qubits $\phi_\mathrm{nopt}(v)$. It suggests that this type of instance is less favorable to the quantum annealer and will be more easily solved by classical computers.

We evaluate the efficiency of our method using a complete graph of size $n=100$. The CME method finds the mapping function of the complete graph $G_\mathrm{s}$ to D-Wave \textit{Advantage6.4}, using 982 physical qubits. The method described in the previous section is used to generate source graphs $G_\mathrm{s}'$ of diverse densities $d \in \{0.9, 0.8, ..., 0.1\}$ with their associated near-optimal mapping $\phi_\mathrm{nopt}$. For each source graph $G_\mathrm{s}'$, the CMR method is run 100 times with a time limit set to 5 minutes to find a second mapping $\phi_\mathrm{100\_CMR}$ that uses the least number of nodes in $G_\mathrm{t\_QPU}$. \Cref{fig:opt_emb_vs_mm_emb}a compares the values of $r_\mathrm{o}$ for mappings: $\phi_\mathrm{nopt}$ and $\phi_\mathrm{100\_CMR}$ and clearly shows that $\phi_\mathrm{nopt}$ mappings use a number of nodes close to the optimal bound defined in \cref{eqn:opt_phys_qubits}. The mappings $\phi_\mathrm{100\_CMR}$ use more nodes that $\phi_\mathrm{nopt}$ for every instances. The worst case behavior of $\phi_\mathrm{100\_CMR}$ is on sparse instances, whereas $\phi_\mathrm{nopt}$ produces embeddings that can be considered near-optimal. We then use the graphs $G_\mathrm{s}'$ to create max-cut instances and compute the ratio of cut sizes obtained using each mapping $\phi_\mathrm{nopt}$ and $\phi_\mathrm{100\_CMR}$ with D-Wave \textit{Advantage6.4}. \Cref{fig:opt_emb_vs_mm_emb}b shows that the best-cut size is increased when using the mapping $\phi_\mathrm{nopt}$ instead of the mapping $\phi_\mathrm{100\_CMR}$. It confirms that the mappings found by our method, in addition to using fewer qubits, increase the performance of the D-Wave \textit{Advantage6.4}.

\subsection{Instance Set}

In the rest of the paper, we generate instances from a complete graph of size $n=174$. It is the largest clique that can be embedded with the CME method on \textit{Advantage6.4}, using $2918$ physical qubits, representing approximately $52\%$ of the physical working qubits of the chip. Using the method described above, we generate graphs of various densities $G_\mathrm{s}'$ with their associated near-optimal transformation $\phi_\mathrm{nopt}$ on \textit{Advantage6.4} (see \cref{table:embedding} for instances description). The source graphs $G_\mathrm{s}'$ are then used to create instances of three different optimization problems with 30 instances for each density. The two first problems are unweighted and weighted max-cut problems defined as:
\begin{figure}[t]
    \centerline{\includegraphics[page=1,width=\columnwidth]{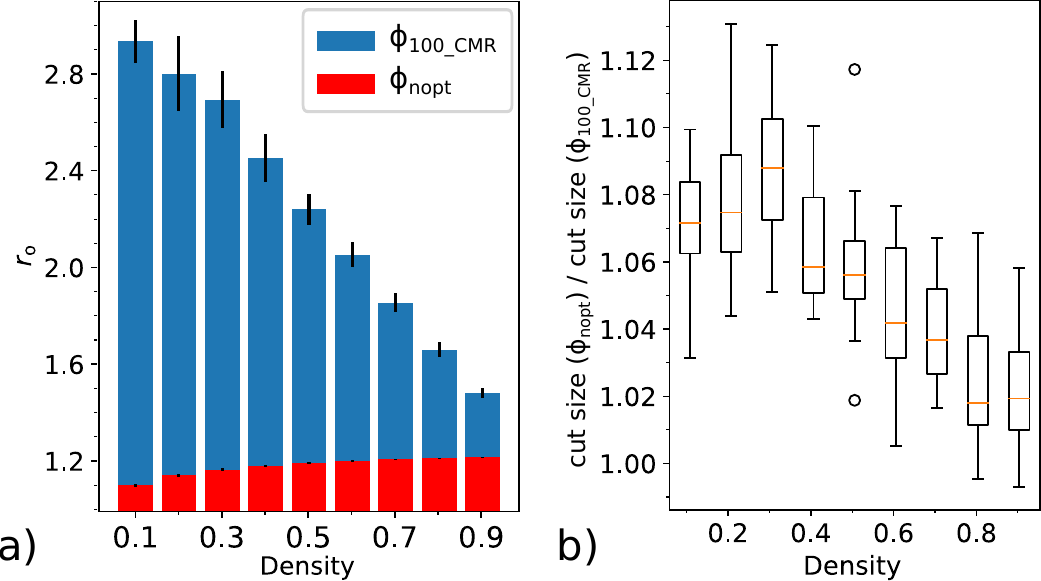}}
    \caption{Performance comparison of mapping functions found by our method $\phi_\mathrm{nopt}$ and the best among 100 tries of CMR method $\phi_\mathrm{100\_CMR}$. 30 instances are generated for each density. a) Comparison using the overhead ratio (see \cref{eqn:overhead_ratio}). Error bars show the standard deviations b) Ratio of the best cut size obtained for each mapping for each instance.}
    \label{fig:opt_emb_vs_mm_emb}
\end{figure}

\begin{equation}
    \min C(\mathbf{s}) = \sum_{(u, v) \in E_\mathrm{s}'} J_{uv} s_u s_v
\end{equation}
where $s_u, s_v \in \{-1, +1\}$. $J_{uv} = 1$ for unweighted max-cut and $J_{uv} \in \{-\frac{128}{128}, ..., -\frac{1}{128}, \frac{1}{128}, ..., \frac{128}{128}\}$ for weighted max-cut. Weights are used to challenge the precision and long-range coherence of the QPU. The precision is the same as in \cite{king2024computational}. The third problem is a weighted Maximum Independent Set (MIS) defined by the cost function:
\begin{equation}
    \min C(\mathbf{x}) = -\sum_{v \in V_\mathrm{s}'} \omega_v x_v + \sum_{(u, v) \in E_\mathrm{s}'} \omega_{uv} x_u x_v
\end{equation}
where $x_u, x_v \in \{0, 1\}$, $\omega_v  \in \{\frac{1}{128}, ..., \frac{128}{128}\}$ and $\omega_{uv} = 2$. $\omega_{uv}$ acts as the penalty term used to enforce the hard constraints for D-Wave and Tabu search. The constraints are encoded with inequalities on the solver Gurobi to reduce the search space.

\subsection{Solver Settings}

\subsubsection{\textit{Advantage6.4} settings} 
\begin{figure}[b]
    \centerline{\includegraphics[page=1,width=0.98\columnwidth]{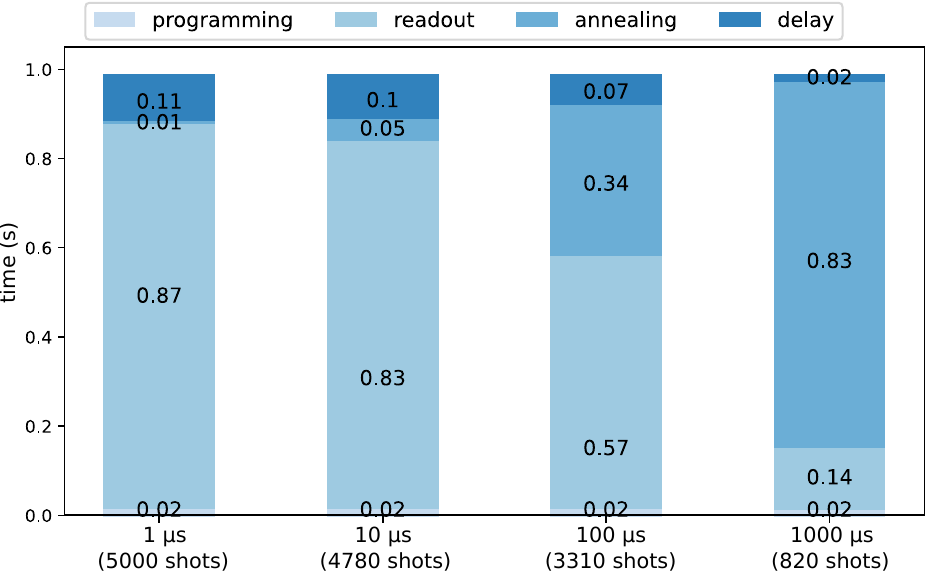}}
    \caption{QPU access time repartition for different annealing times per shot with a total running time limit of $1s$. Each color represents the fraction of time used for programming the QPU, annealing the $n$ shots, reading the results of the $n$ shots, and delays between the $n$ shots.}
    \label{fig:time_repartition}
\end{figure}

\begin{figure}[t]
    \centerline{\includegraphics[page=1,width=\columnwidth]{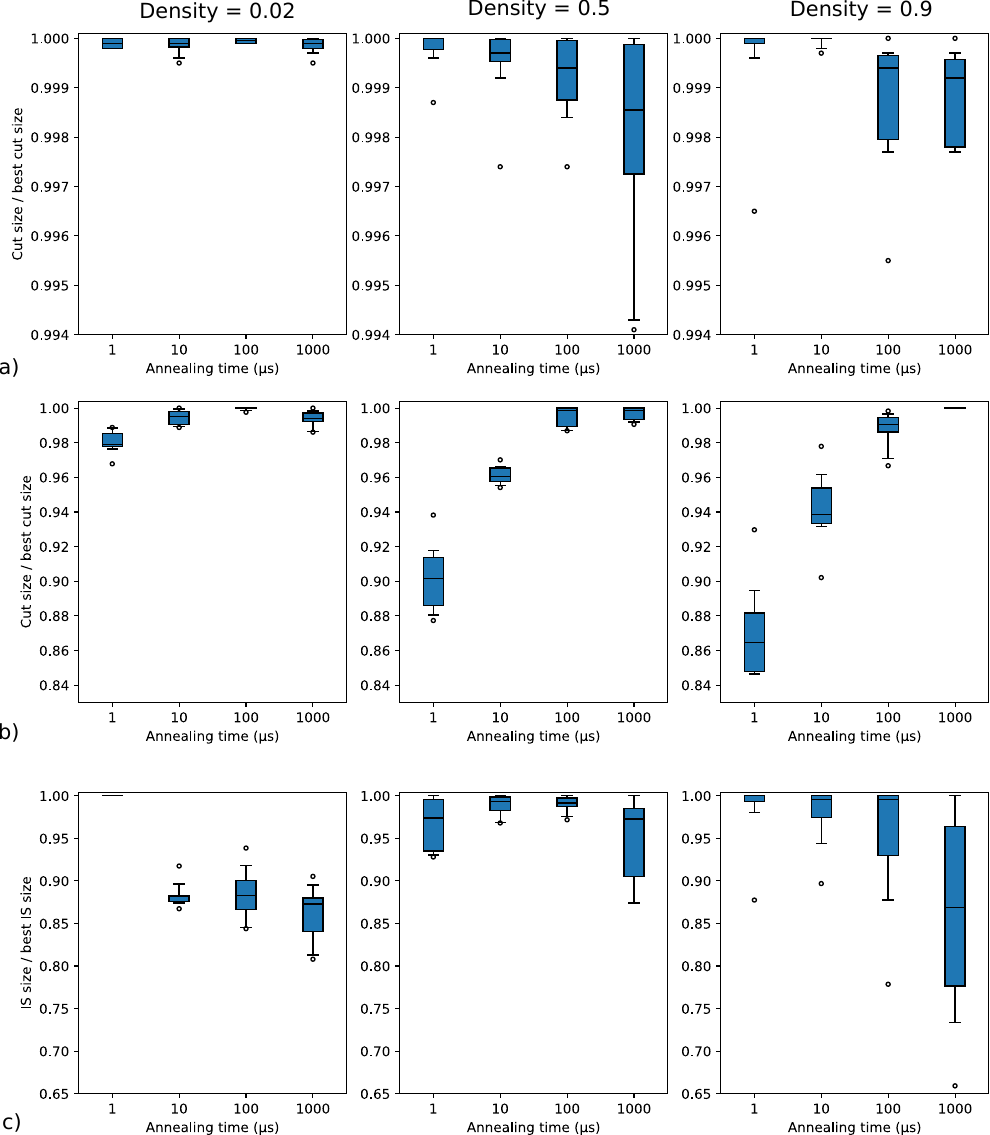}}
    \caption{Annealing time scan with a quantum processing time limit of $1s$ for three types of problems with various densities. a) Unweighted maxcut instances b) Weighted maxcut instances c) Weighted MIS.}
    \label{fig:an_time_scan}
\end{figure}

The performance of the quantum annealer is evaluated using an access time limit of 1 second. Each instance is mapped on the QPU using the mapping $\phi_\mathrm{nopt}$. The repartition of the processing time considering different annealing times $\{1, 10, 100, 1000\} \mu s$ with respectively $\{5000, 4780, 3310, 820\}$ shots is illustrated in \cref{fig:time_repartition}. It shows the existing trade-offs between the number of shots and the duration of the annealing time per shot. \Cref{fig:an_time_scan} shows the performance of D-Wave considering different values of annealing time for each problem. The average optimal annealing time seems to be problem-dependent. Considering these results, we choose $1\mu s$ and $5000$ shots for unweighted max-cut, $1000 \mu s$ and $820$ shots for weighted max-cut and $1 \mu s $ and $5000$ shots for weighted MIS problems. The \textit{uniform\_torque\_compensation} method implemented on D-Wave sets the chain strength used to maintain ferromagnetic couplings between the physical qubits $\phi_\mathrm{nopt}(v)$. The weight associated with a single qubit $h_v$ is uniformly spread over the physical qubits $\phi_\mathrm{nopt}(v)$. The same method is used for edges. A majority vote is used to unembed the problem. We do not use annealing offset or spin reversal methods. MIS instances are post-processed to avoid constraint violations. We iteratively and randomly select edges that violate the constraint and only keep the node with the higher weight in the independent set. The benchmark is done using the best solution found among all shots for each instance. 
\subsubsection{Tabu Search settings} The Tabu search is based on the method introduced in \cite{Palubeckis2004}. The Tabu search is run directly on the instance associated with the source graph $G_\mathrm{s}'$. The processing time limit is set to $1 s$ and the tenure length is set to $|V_\mathrm{s}'|/4$. The Tabu search uses a single core of processor \textit{AMD EPYC 7702P}. MIS instances are post-processed with the same method used for D-Wave.

\begin{table}[t]
    \centering
    \caption{Properties of $G_\mathrm{s}'$ graphs used to build optimization problem instances. These graphs are mapped on 2918 qubits.}
    \begin{tabular}{|c|c|c|||c|c|c|}
        \hline
        Density & Avg $|V_\mathrm{s}'|$ & $r_\mathrm{o}$ & Density & Avg $|V_\mathrm{s}'|$ & $r_\mathrm{o}$\\
        \hline
        0.02 & 1318 & 1.06 & 0.1 & 565 & 1.14 \\
        0.03 & 1062 & 1.08 & 0.2 & 395 & 1.16 \\
        0.04 & 912 & 1.10 &  0.3 & 321 & 1.17 \\
        0.05 & 810 & 1.11 & 0.4 & 277 & 1.18 \\
        0.06 & 737 & 1.12 & 0.5 & 248 & 1.19 \\
        0.07 & 680 & 1.12 & 0.6 & 226 & 1.19 \\
        0.08 & 635 & 1.13 & 0.7 & 209 & 1.19 \\
        0.09 & 597 & 1.13 & 0.8 & 195 & 1.20 \\
             &     & & 0.9 & 184 & 1.20 \\
        \hline
    \end{tabular}
    \label{table:embedding}
\end{table}

\subsubsection{Gurobi settings} Gurobi \cite{gurobi} is run with a time limit of $60s$. It is used to establish a reference solution with which the Tabu search and D-Wave are compared. The solver is set with default parameters. The solver is run on the instance associated with the source graph $G_\mathrm{s}'$. For the MIS problem, the constraints are directly encoded with inequalities. Gurobi is parallelized on 20 cores of a processor \textit{AMD EPYC 7702P}.

\subsubsection{Random solver settings} A random solver is used on MIS instances. It iteratively and randomly selects a node in the set $V_\mathrm{s}'$ and removes from $V_\mathrm{s}'$ all the neighbors of this node. This method generates the same number of solutions generated by D-Wave, i.e., 5000 solutions. The best MIS is then selected among these shots. We did not use this method for the max-cut problem as random solution cuts are far from being competitive with D-Wave and Tabu search solutions.

\section{Results and Discussion}\label{section:4}

\begin{figure*}[t]
\centerline{\includegraphics[page=1,width=\textwidth]{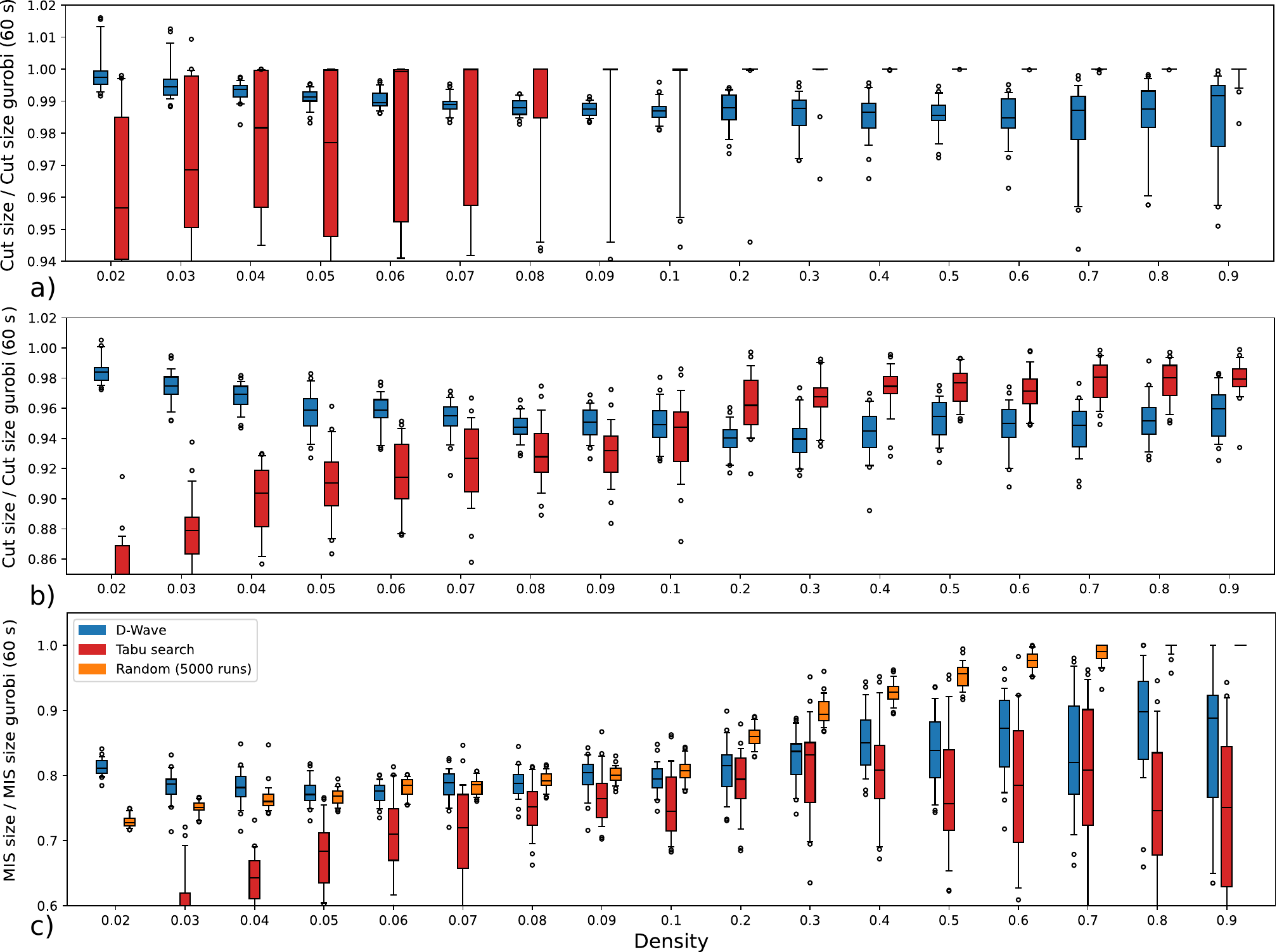}}
    \caption{Benchmark of D-Wave \textit{Advantage6.4} and Tabu search (1s time limit). The results are expressed as ratio of the best result found by Gurobi (60s runtime limit). The size of each instance is provided in \cref{table:embedding} a) Unweighted max-cut b) Weighted max-cut c) MIS problem with a random solver.}
    \label{fig:dwave_benchmark}
\end{figure*}

We compare the performance of D-Wave \textit{Advantage6.4} against classical solvers to approximate solutions of optimization problems introduced in the previous section. \Cref{table:embedding} shows the average number of logical variables for each density processed in \cref{fig:dwave_benchmark}. We recall that the density of the graph is inversely proportional to the number of variables in the graph $G_s'$. Hence, the size of instances increases when the density decreases. The results of D-Wave and the Tabu search are expressed as a ratio of a reference solution computed with Gurobi with a runtime limit set to $60s$. In general, D-Wave performs well with respect to Tabu search on sparse instances under 0.1 density. Under this density, the logical problem has more than 565 variables, which limits the performance of the Tabu search due to the large size of the solution space. We observed that the Tabu search did not have sufficient processing time for large instances, leading to almost random solutions to these instances. It explains why the box plots are stretched for Tabu search for small densities. The performance transition between the quantum and classical heuristics occurs at density $d=0.06$ for unweighted max-cut, around $d=0.1$ for weighted max-cut instances and at $d=0.05$ for MIS instances. On very sparse instances ($d<0.03$), the \textit{Advantage6.4} is able to outperform Gurobi on few instances of unweighted and weighted max-cut, showing that the QPU may be useful against classical methods for very sparse instances having more than 1000 variables. For denser instances, lots of extra qubits are used to encode the problem on the QPU. These instances become easier for classical solvers as the search space reduces exponentially with the density but stays approximately the same for the QPU as the number of physical qubits remains the same due to the embedding. In addition, long chains of physical qubits require strong ferromagnetic couplings to maintain spin orders in chains $\phi_\mathrm{nopt}(v)$. It adds extra energy to the system and impacts the couplers' precision due to the rescaling of Hamiltonian weights. Both Tabu search and D-Wave QPU do not seem to be adapted to approximate solutions to the MIS problem. These methods are unable to compete with the random solver used to generate valid maximal independent sets. D-Wave only starts to be competitive on instances with densities lower than 0.02. This behavior is reminiscent of the fact that both D-Wave and Tabu search performances are limited by the constraints encoding using penalty methods and cannot produce valid solutions in a dense regime. 

Several points have to be clarified and considered carefully to avoid misleading conclusions. The aim of this benchmark is not to study the scaling of the QPU but rather to determine when the quantum computer performs best at solving instances that use the same physical requirements (in our case, the number of physical qubits). The settings of this experiment only involved tuning the annealing time, which has a huge impact when working on time-limited tasks. We did not want to include other complex parameter optimization such as the advanced tuning of the chain strength \cite{gilbert2024quantum, willsch2022benchmarking, Le2023}, annealing schedules \cite{Adame2020, Khezri2022} or advanced calibration methods called shimming \cite{chern2023tutorial}. The main idea of our study is to provide an objective view of D-Wave's average performance without employing complex preprocessing methods. Nevertheless, the study of the trade-offs brought by these methods represents a relevant perspective for future work. The comparison with classical heuristics can also be enhanced by using algorithms that excel on sparse instances of the maxcut problem \cite{Rehfeldt2023}.

The minor-embedding task is a step that is usually excluded from quantum benchmark studies, which almost only consider the pure annealing time without considering delays, readouts and programming time. However, a huge amount of classical preprocessing time can be spent to optimize the mapping function, making comparing existing studies difficult. Instead, our approach permits the evaluation of the QPU on near-optimally mapped instances, giving insights into the properties that leverage the power of QA.

To conclude the discussion, this benchmark methodology is not fair and representative of the performance of D-Wave systems' ability to solve real-world instances. It rather acts as a performance evaluation of D-Wave systems' ability to solve large optimization instances based on an ideal mapping of the instance on the QPU, using QA default settings. This methodology could be used to detect optimization problems that could benefit from a quantum advantage.

\section{Conclusion}

This article introduced a new method to create source graphs of various densities near-optimally mapped on a QA. This set of graphs was then used to generate large optimization problem instances to benchmark D-Wave QPU against classical solvers. Considering an equal processing time limit, the QPU was competitive with Tabu search for instances with densities inferior to $0.1$ on the Weighted max-cut problem. It even outperformed Gurobi solver on some really sparse unweighted max-cut instances, whereas Gurobi had $60$ times the processing time of D-Wave QPU. Conversely, the QPU seemed less efficient at approximating similar-sized constrained problems. It is important to remember that this set of \textit{crafted instances} is very favorable and specifically designed for D-Wave QA. Our results can be seen as the benchmark of the default behavior of D-Wave QPU on large instances of ideal shape. On real instances, the mapping function found by embedding methods such as CMR will not be as efficient as the one designed in this paper.

A relevant perspective is to study other optimization problems using this methodology to build the instance set. It could also be used to evaluate the benefits of advanced preprocessing methods of quantum annealers. Finally, this methodology can also be adapted to benchmark variational quantum algorithms working on gate-based quantum computers, considering the generation of instances adapted to the topology of the quantum chip with a planted optimal swapping network. 

\bibliographystyle{ieeetr}
\bibliography{bibliography}

\end{document}